# Evaluative Assessment of an X-band Microstrip Patch Antenna for Wireless Systems

A. F. Salami[1*], O. S. Zakariyya[2], B. O. Sadiq[3] and O. A. Abdulrahman[4]

[1]Department of Computer Engineering, University of Ilorin, Ilorin, Kwara State, Nigeria.
[2]Department of Electrical and Electronics Engineering, University of Ilorin, Ilorin, Kwara State, Nigeria.
[3,4]Department of Electrical and Computer Engineering, Ahmadu Bello University, Zaria, Kaduna State, Nigeria.
[1]salami.af@unilorin.edu.ng, [2]zakariyya.os@unilorin.edu.ng, [3]bosadiq@abu.edu.ng, [4]olaniyanabdulrahman@gmail.com

**ABSTRACT**

Microstrip patch antennas (MPAs) are rapidly gaining more attention due to the proliferation of communication devices and systems with frequencies becoming more suitable for the size and performance of this type of antenna. Due to recent advancements in semiconductor technology, high dielectric constant materials are used to achieve additional size reduction which has made MPAs very useful and popular in the design of mobile devices and wireless systems. However, MPAs suffer from problems associated with narrow bandwidth and low gain. Techniques employed for improving the performance of MPA hinge on tweaking features such as the patch size, substrate height, ground plane size and feeding method. In view of this, this research designs and analyzes the performance of an X-band MPA for wireless systems using CST Microwave Studio. Including the ground plane, the proposed design has a low profile structure of 17 mm × 17 mm × 1.6 mm which is suitable for wireless systems. The proposed design also resonates at a frequency of 10 GHz with an omnidirectional radiation pattern exhibiting a gain of 7.2 dBi. Return Loss, VSWR, Gain and Radiation Pattern are the performance indicators employed in this research. The proposed MPA design demonstrates marked performance improvement when benchmarked with a similar MPA designed for 5G applications.

**Keywords:** Patch Antenna, Microstrip Line, Wireless Systems, Mobile Applications, Communication Networks, Antenna Parameters

**1.0 INTRODUCTION**

Antennas are essentially metallic structures acting as transducers with a region of transmission between the guided wave structures and free space, or vice versa. This guiding wave structure is usually in form of a waveguide or two-wire transmission line leading from a transmitter or receiver to the antenna. Generally, antennas are good conductors efficiently and effectively designed with dimensions, shapes and sizes suitable for the desired transmission and reception of electromagnetic energy [1, 5].

Microstrip is basically a transmission line fabricated with PCB technology and effectively designed for the transmission and reception of microwave frequency signals [1, 3]. It consists of a conducting strip with a substrate (dielectric layer) separating the microstrip from the ground plane [4, 6]. This means that the antenna and other microwave components are formed from the microstrip with the entire circuitry or device existing as a metallization pattern on the substrate. Microstrip is relatively lighter, more compact and less expensive when compared to the conventional waveguide technology [3, 4].

The patch is usually made of a conducting material (gold or copper) and it can be designed to assume specific shapes (square, rectangular, triangular, dipolar, circular, circular ring, elliptical, disc sector, ring sector) [1, 3]. The patch acts as the radiator and it is etched together with the feed lines on the substrate using photolithographic technology [4, 20]. This research focuses on the rectangular patch. The length (*L*) of the rectangular patch is usually in the range of $0.3333\lambda_o < L < 0.5\lambda_o$, where $\lambda_o$ is the free-space wavelength [2, 19]. The patch thickness (*t*) is also usually very thin which means that $t \ll \lambda_o$. The substrate height (*h*) is typically in the range of $0.003\lambda_o \leq h \leq 0.05\lambda_o$. The substrate dielectric constant ($\varepsilon_r$) is normally in the range of $2.2 < \varepsilon_r < 12$ [2, 19, 23].

**2.0 MECHANISM OF OPERATION**

The basic form of the patch antenna is a flat plate on a PCB or ground plane. In order to couple EM energy into and out of the patch, a feed probe placed at the center of the coax conductor is used to serve this purpose [1, 4, 19]. The instantaneous phase of the applied signal continuously changes the location of the E-field maximum (positive) and minimum (negative). At the patch center, the E-field is normally zero while the maximum and minimum values are detected at the opposite sides of the patch's periphery [19, 20]. However, the E-field does not terminate abruptly at the peripheries, but rather the E-field extends beyond the peripheries to a certain degree. This phenomenon (leaky cavity concept or open-circuit boundary condition) leads to the fringing fields which are the physics behind the radiating patch [19, 23].

Based on the leaky cavity theory, the mode for analyzing patch antennas can be written as $TM_{nmz}$, where TM stands for Transversal Magnetic field distribution [6, 19, 20]. This implies that only three field components are considered instead of the six fields treated in conventional analysis. These three field components are: (1) E-field in the *z*-direction (*z*), (2) H-field in the *x*- direction or resonance length direction (*n*), and (3) H-field in the *y*-direction or impedance width direction (*m*) [1, 2, 23]. The *z*-component is usually ignored due to the fact that E-field variation in the *z*-direction is negligible. The H-field variation in the *y*-direction is negligible, hence *m* is zero. The field has one minimum to maximum variation in the *x*-direction, hence *n* is one for the case of the basic rectangular patch. Therefore, the fundamental mode for a rectangular patch is typically written as TM$_{10}$ [19, 20, 23].

In addition to this, it should be noted that the patch can be further simplified by approximating it to a cavity with its top and bottom surfaces as Perfect Electric Conductors (PEC) and Perfect Magnetic Conductors (PMC) by the sides [1, 2, 6]. Inside the cavity, the E-field is in the *z*-direction but independent of the *z*-coordinate. This means that the patch cavity mode is double-indexed (*n, m*). The analytical process for the rectangular patch necessitates constant current density.

The radiation from the *x-y* patch with constant current density $M_s$ is computed from the electric vector potential, *F*. For simplicity, let ***i*** = *x*-direction, ***j*** = *y*-direction and ***k*** = *z*-direction. The vector potential will only have a *y*-component (*F* = $F_y$***j***) since $M_s$ has only *y*-component [1, 5, 6]. The resultant vector potential, $F_y(r, \theta, \varphi)$ is expressed as [5, 6]:

$$F_y(r, \theta, \emptyset) = \frac{\epsilon}{4\pi} \int_{\frac{-h}{2}}^{\frac{h}{2}} \int_{\frac{-W}{2}}^{\frac{W}{2}} \frac{M_y}{r_{PQ}} e^{-jk_0 r_{PQ}} dx'dy' \ldots \ldots \ldots \ldots \ldots \ldots (1)$$

With reference to (1), $\varepsilon$ is the permittivity, *h* is the patch length, *W* is the patch width, $M_y = -2E_0$, $E_0$ is the phasor of the E-field at the radiating patch and $r_{PQ} = r - x'\sin(\theta)\cos(\varphi) - y'\sin(\theta)\sin(\varphi)$ [1, 5, 6]. The final expression for the vector potential is given as [5, 6]:

$$F_y = \frac{-\epsilon E_0 W h}{2\pi r} e^{-jk_0 r} \frac{\sin X}{X} \frac{\sin Y}{Y} \ldots \ldots \ldots \ldots \ldots \ldots \ldots (2)$$

With respect to (2), $X = [(k_0 h)/2]\sin(\theta)\cos(\varphi)$, $Y = [(k_0 W)/2]\sin(\theta)\sin(\varphi)$, *r* is the far-field distance, $k_0 = \omega\eta\varepsilon_0$, and $\eta = (\mu_0/\varepsilon_0)^{1/2}$ is the intrinsic impedance [1, 5, 6]. Based on the relationship between the vector potential and the E-field, the far-field solutions for the E-field have the following results [5, 6]:

$$E_r \approx 0 \ldots \ldots \ldots \ldots \ldots \ldots \ldots \ldots \ldots \ldots \ldots \ldots \ldots \ldots \ldots \ldots \ldots (3)$$

$$E_\emptyset = j\omega\eta F_\theta = jk_0 W \frac{V_0}{2\pi r} e^{-jk_0 r} \left(\cos\theta\sin\phi \frac{\sin X}{X} \frac{\sin Y}{Y}\right) \ldots \ldots \ldots (4)$$

$$E_\theta = -j\omega\eta F_\phi = -jk_0 W \frac{V_0}{2\pi r} e^{-jk_0 r} \left(\cos\phi \frac{\sin X}{X} \frac{\sin Y}{Y}\right) \ldots \ldots \ldots (5)$$

From (4) and (5), $V_0 = hE_0$, $F_\theta = F_y \cos(\theta)\sin(\varphi)$ and $F_\varphi = F_y \cos(\varphi)$ [5, 6]. The two peripheries of the patch is equivalent to an antenna array of two elements with an inter-element distance of *L* =

λ/2. The normal components of the E-field at the two peripheries along the width are in opposite direction and thus out of phase. This means these components cancel each other out in broadside direction. However, the tangential components are in phase which makes the resulting fields add up to give maximum radiated field normal to the surface [1, 5, 6]. Hence the two peripheries can be represented as two elements (λ/2 apart and excited in phase) radiating in the half space above the PCB or ground plane. The array factor (*AF*) is therefore expressed as [5, 6]:

$$AF_{12} = 2\cos\left(\frac{k_0 L_{eff}}{2}\cos\theta\right) \quad\quad\quad\quad\quad\quad (6)$$

From (6), $L_{eff} = L + 2\Delta L$ is the effective patch length [1]. Therefore, the total radiated E-field is obtained as [5, 6]:

$$E_\phi^t = \frac{jk_0 W V_0}{\pi r} e^{-jk_0 r}\left(\cos\theta\sin\phi \frac{\sin X}{X}\frac{\sin Y}{Y}\right) \times \left[\cos\left(\frac{k_0 L_{eff}}{2}\cos\theta\right)\right] \quad (7)$$

$$E_\theta^t = \frac{-jk_0 W V_0}{\pi r} e^{-jk_0 r}\left(\cos\phi \frac{\sin X}{X}\frac{\sin Y}{Y}\right) \times \left[\cos\left(\frac{k_0 L_{eff}}{2}\cos\theta\right)\right] \quad (8)$$

By letting $Z = (k_0 L_{eff}/2)cos(\theta)$, the field pattern of the patch is expressed as [5, 6]:

$$f(\theta,\phi) = \sqrt{\bar{E}_\phi^2 + \bar{E}_\theta^2} = \sqrt{1 - \sin^2\phi\sin^2\theta}\,\frac{\sin X}{X}\frac{\sin Y}{Y}\cos Z \quad\quad (9)$$

For the purpose of simple illustration, the E-plane pattern (*x-z* plane, φ = 0°, 0°≤θ≤180°) and H-plane pattern (x-y plane, θ = 90°, 0°≤φ≤90°, 270°≤φ≤360°) will yield the following simplified expressions [5, 6]:

$$f_E(\theta) = \frac{\sin\left(\frac{k_o h}{2}\sin\theta\right)}{\frac{k_o h}{2}\sin\theta}\cos\left(\frac{k_0 L_{eff}}{2}\cos\theta\right) \quad\quad\quad\quad (10)$$

$$f_H(\phi) = \cos\phi\,\frac{\sin\left(\frac{k_o h}{2}\cos\phi\right)}{\frac{k_o h}{2}\cos\phi}\frac{\sin\left(\frac{k_o W}{2}\sin\phi\right)}{\frac{k_o W}{2}\sin\phi} \quad\quad\quad\quad (11)$$

Due to the structure of the surface, the current will have its maximum value at the center of the patch while the E-field will have its maximum values at the two radiating peripheries [1, 6]. The patch width is normally chosen to be larger than the patch length as the bandwidth is proportional to the patch width (*W*). On the other hand, the radiation quality factor (*Q*) is inversely proportional to the substrate thickness (*h*). This means that the strength of radiated field and resonant input resistance are almost independent of the substrate thickness (*h*), provided losses are ignored. This phenomenon gives the physical reason why the patch antenna radiates effectively even with a very thin substrate [2, 19, 23].

As a result of the leaky cavity phenomenon, an effective dieletric constant ($\varepsilon_{eff}$) is introduced to account for the fringing effects. The value of $\varepsilon_{eff}$ is a little bit less than the substrate dielectric constant ($\varepsilon_r$) because the fringing fields along the peripheries are not totally contained in the substrate but portion of the fields are dissipated into the air [1, 19, 20]. The effective dielectric constant is expressed as [1]:

$$\varepsilon_{eff} = \frac{\varepsilon_r + 1}{2} + \frac{\varepsilon_r - 1}{2}\left[1 + 12\frac{h}{W}\right]^{\frac{-1}{2}} \quad \ldots (12)$$

In the $TM_{10}$ mode, the patch length is slightly less than $\lambda/2$, where $\lambda = \lambda o/\sqrt{\varepsilon_{eff}}$ is the wavelength in the dielectric medium, $\lambda o$ is the free-space wavelength and $c$ is the speed of light. In the $TM_{10}$ mode, the field varies one $\lambda/2$ cycle along the length with no difference along the width of the patch [1, 6, 19]. The dimension of the patch is now being extended due to this fringing effect by $\Delta L$ which is expressed as [1]:

$$\Delta L = 0.412h\frac{(\varepsilon_{eff} + 0.3)\left(\frac{W}{h} + 0.264\right)}{(\varepsilon_{eff} - 0.258)\left(\frac{W}{h} + 0.8\right)} \quad \ldots (13)$$

The resonant frequency for any $TM_{mn}$ mode rectangular MPA is given as [1]:

$$f_0 = \frac{c}{2\sqrt{\varepsilon_{eff}}}\left[\left(\frac{m}{L}\right)^2 + \left(\frac{n}{W}\right)^2\right]^{\frac{1}{2}} \quad \ldots (14)$$

For a given resonant frequency $f_0$, the effective patch length ($L_{eff}$) and patch width ($W$) can also be expressed as [1]:

$$L_{eff} = \frac{c}{2f_0\sqrt{\varepsilon_{eff}}} \quad \ldots (15)$$

$$W = \frac{c}{2f_0\sqrt{\frac{\varepsilon_r + 1}{2}}} \quad \ldots (16)$$

## 3.0 SELECTED SCHEMES FOR BANDWIDTH ENHANCEMENT

Researchers have proffered various means of solving the bandwidth issues of rectangular MPAs. One of such ways is to increase the height of the dielectric substrate or decrease the dielectric constant. However, increasing the substrate height defeats the designer's aim of building a compact low-profile antenna structure while decreasing the dielectric constant will result in excessively wide lines that will make matching the circuit to the patch impractical [2, 13]. For clarity and better understanding, the mathematical relationship linking the bandwidth (*B*), wavelength ($\lambda$), height of the dielectric (*h*) and dielectric constant ($\varepsilon_r$) is expressed as [1, 6]:

$$B = 3.77 \left(\frac{\varepsilon_r - 1}{\varepsilon_r^2}\right) \frac{h}{\lambda} \ldots \ldots \ldots \ldots \ldots \ldots \ldots \ldots \ldots \ldots \ldots \ldots (17)$$

However, it must be mentioned that (17) is only valid when $h/\lambda \ll 1$ [1, 6]. This bandwidth is simply the fractional bandwidth which is relative to the center frequency for a specified voltage standing wave ratio (VSWR) less than 2:1, where VSWR = $(1+|\Gamma|)/(1-|\Gamma|)$, and $\Gamma$ is the reflection coefficient defined as $\Gamma = (Z - Z_0)/(Z + Z_0)$. $Z_0$ is the characteristic impedance normally set at 50Ω [1, 2, 6]. The underlying idea behind constraining the VSWR at 2:1 is to have a reflection coefficient at -10dB for the design [2, 13].

Apart from the numerical-based improvement technique, another group of researchers proposed the use of U-slot and L-probe designing low-profile MPAs [13, 27]. This approach is basically a size reduction technique which explores options of using microwave substrate material and including short-out walls and/or pins in the designs. By incorporating these adjustments, considerable improvements were observed with respect to the solely numerical-based approaches. Another technique proposed by researchers having similar bandwidth improvement is the use of electromagnetic band-gap (EBG) structures with different shapes and sizes [10, 26].

In addition to this, the use of compound techniques was suggested and implemented by some researchers as an alternative solution [28]. This technique entails adjusting the displacement of two patches by setting two pairs of conducting bars around the lower patch (parasitic radiator) while loading a capacitive disk on top of the probe. This results in a kind of stacked MPA with slight bandwidth improvements when compared to the slot/probe-based and EBG techniques [24, 26, 28].

Another group of researchers utilized filter design techniques as an impedance matching mechanism to improve the bandwidth of an aperture-coupled MPA [29]. However, this approach did not yield much marked improvements in bandwidth due to the complexities resulting from the feed configuration. In order to tackle this challenge, the use of unbalanced structures in the design of MPA for better bandwidth and VSWR performance was proposed [30, 31]. This is hybrid approach incorporates mathematical analysis into the practical design of the patch antenna by employing full-wave spatial domain method of moments together with closed-form Green functions to analyze and characterize microstrip discontinuities. The obtained numerical results demonstrated good agreement with the practical measurement data with respect to marked improvement in bandwidth performance [30, 31].

Other methods have been proposed for improving the bandwidth and reducing the size of MPA, namely; using high dielectric material, utilizing shorting techniques and increasing the patch antenna length via shape optimization techniques [3, 21, 22]. In addition to this, researchers have shown that significant improvements can be achieved with respect to bandwidth enhancement and size reduction by loading the peripheries of the patch with inductive elements and inserting capacitive element into the patch [32].

Presently, a lot of MPA designs have been proposed for wireless systems and mobile applications. MPAs and PIFAs are suitable antenna choices for some wireless applications [14, 16, 18]. In one of such approaches, a low profile MPA was proposed in which the peripheries of the patch were systematically folded along with the loading slots based on technical specifications [9]. This resulted in noticeable bandwidth improvement and considerable reduction in the patch length. In a similar approach but with another type of antenna, a single-band Planar Inverted-F antenna operating at 28GHz band was used with similar results in terms of bandwidth improvement and patch length reduction [11].

Some researchers have proposed a low profile MPA for next generation mobile devices and wireless systems [7]. With the ground plane or PCB, their proffered patch antenna has a compact dimension of 20mm × 20mm × 1.6mm. This design will serve as a benchmark for this research work. This benchmarked patch antenna provides a gain of 4.46 dBi, omni-directional radiation pattern and resonates at a frequency of 10.15 GHz [7]. Therefore, this research analyzes and enhances the performance of this benchmarked patch antenna design for wireless systems.

## 4.0 MODELING, DESIGN AND SIMULATION

### 4.1 Proposed Antenna Structure and Dimensions

In comparison to the benchmarked patch antenna design, the proposed patch antenna in this research work uses a microstrip line feeding technique but without a slot. The proposed rectangular patch antenna has a dimension of 6mm × 8.6mm as shown in Figure 1. In comparison with the benchmarked patch antenna dimension (10.2mm × 7mm), the proposed dimension in this research is in accordance with the design rule of thumb which requires the relationship between the patch width (*W*) and patch length (*L*) to be $W \approx 1.5L$. One of the contributions of this design is the inclusion of a quarter-wave transmission line linking the patch and the line feed with dimension of 4mm × 0.53mm. The detailed dimensions for the proposed MPA are as given in Table 1.

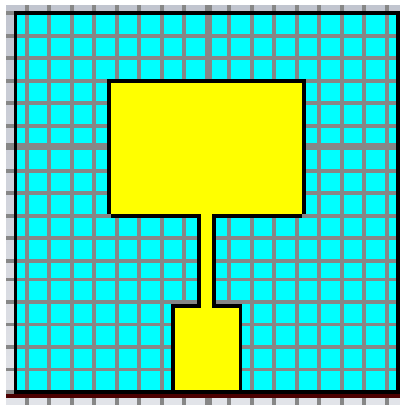

Figure 1: 2D View of Proposed MPA in CST Microwave Studio

In addition to this, the proposed patch antenna design has other parameters such as dielectric constant ($\varepsilon_r = 4.4$), FR-4 lossy substrate material, resonant frequency ($f_0 = 10$ GHz) and substrate thickness ($h = 1.6$ mm). In comparison with the benchmarked patch antenna design, a 50Ω line feed is used to excite the proposed patch antenna. In comparison with the benchmarked patch antenna design (bandwidth of 400 MHz over frequency band of 9.95 GHz – 10.35 GHz), the proposed design has a bandwidth of 500 MHz and covers the frequency band of 9.7542 GHz – 10.25 GHz.

**Table 1: Dimensions of the Proposed MPA**

| PARAMETER | VALUE (mm) |
|---|---|
| Ground Plane Length | 17 |
| Ground Plane Width | 17 |
| Patch Length | 6 |
| Patch Width | 8.6 |
| Substrate Thickness | 1.6 |
| Quarter-wave Length | 4 |
| Quarter-wave Width | 0.53 |
| Feed Line Length | 4 |
| Feed Line Width | 3 |

### 4.2 Simulation of the Proposed MPA

Modeling, design and simulation of the proposed patch antenna is performed using Computer Simulation Technology (CST) Microwave Studio 2014. The simulation procedure entails specifying the antenna parameters, setting the patch dimensions and using the global coordinate system of $U$ (horizontal axis), $V$ (vertical axis) and $W$. $U_{min}$ is the negative $U$ axis, $U_{max}$ is the positive $U$ axis and zero is the origin. $W_p$ is the patch width and $L_p$ is the patch length. In order to model the ground plane, the width of the ground is assumed to be twice the width of patch, i.e., $W_g = 2 \times W_p$. The other major steps in the simulation together with the corresponding settings are shown in Figures 2, 3 and 4.

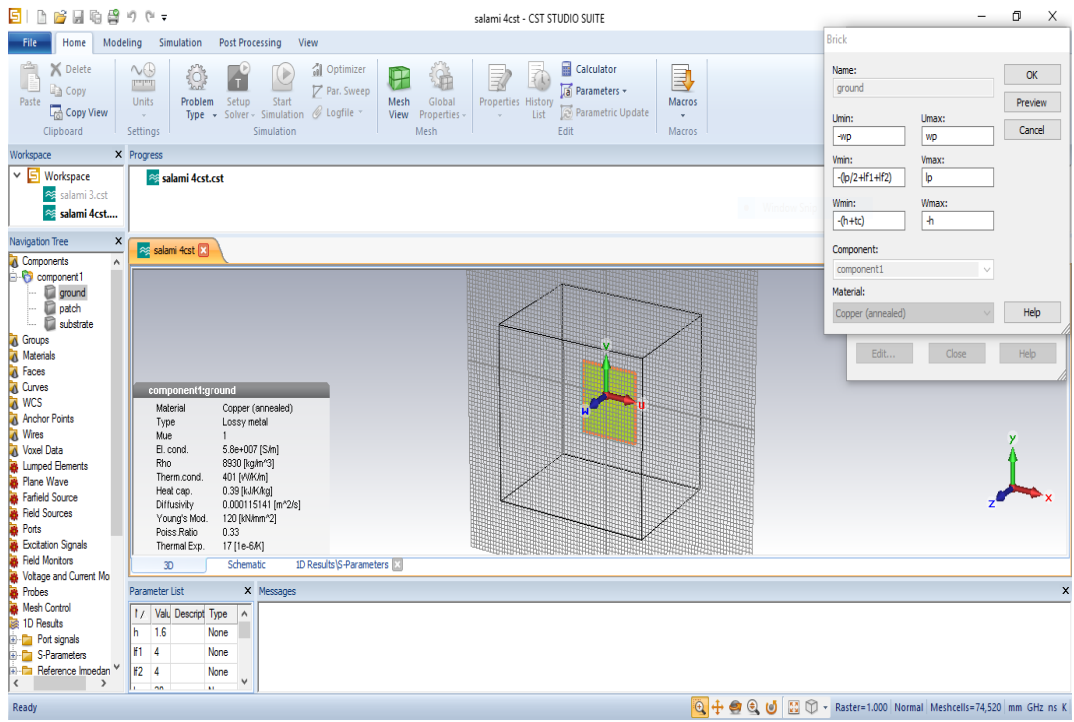

**Figure 2: Modeling of the Ground Plane**

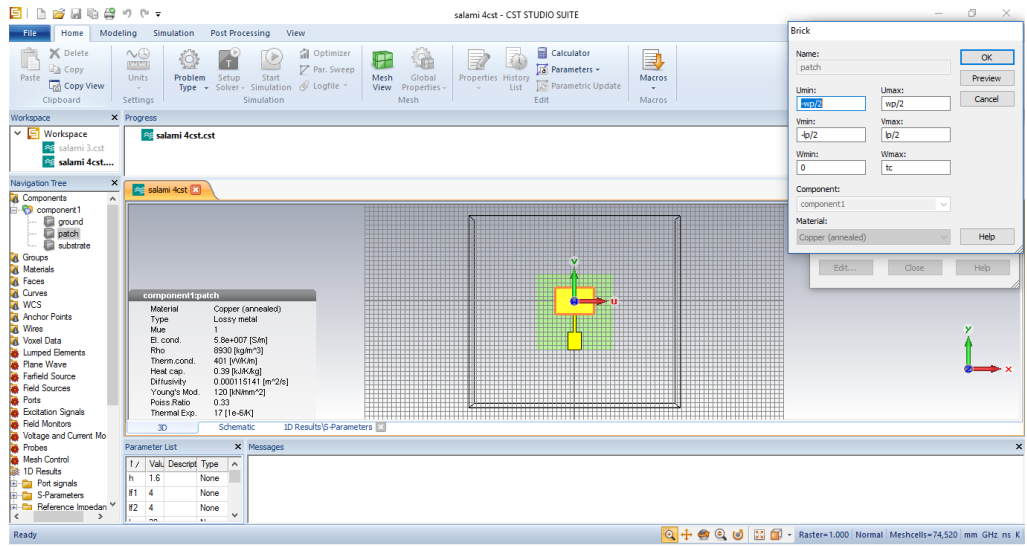

**Figure 3: Modeling of the Microstrip Patch**

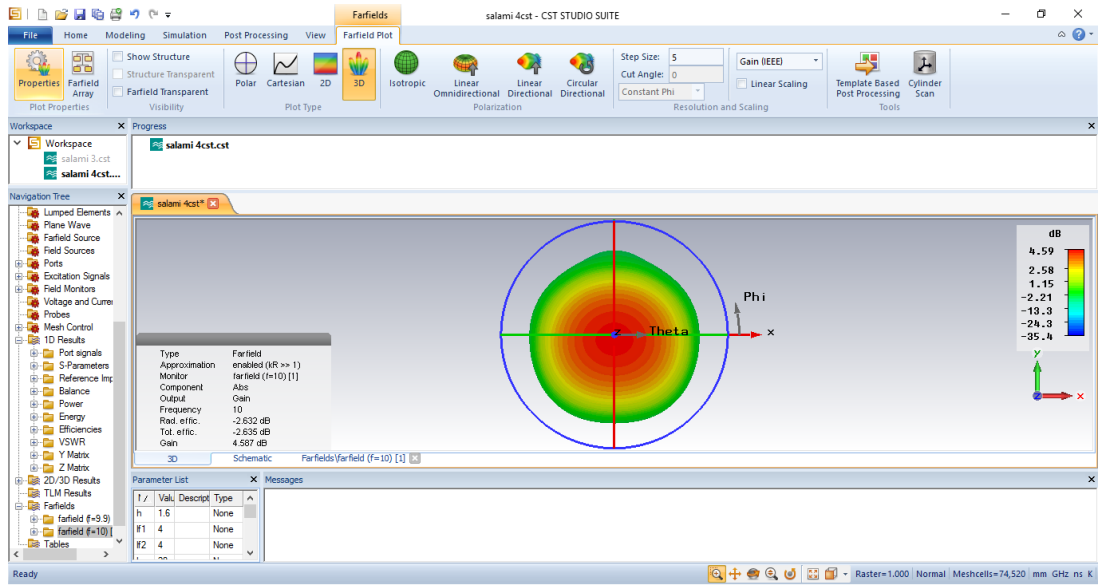

**Figure 4: Obtaining Far-Field Radiation Pattern**

## 5.0 RESULTS AND DISCUSSION

### 5.1 Return Loss

Figure 5 shows the return loss of the proposed MPA. For mobile communication systems, a standard value of -10 dB is acceptable as the baseline for good performance. The proposed MPA has a return loss of -31 dB, resonating at the desired resonant frequency of 10 GHz and covering a frequency band of 9.7542 GHz – 10.25 GHz. The proposed design outperforms the benchmarked patch antenna design in terms of two aspects in this respect. Firstly, the return loss (-31 dB) of the proposed patch antenna design has an improvement of 69.68% over the return loss (-18.27 dB) of the benchmarked patch antenna design. Secondly, the proposed patched antenna is designed with a desired resonant frequency of 10 GHz and after the design, it resonated at the desired resonant frequency. However, the benchmarked patch antenna design was designed with a resonant frequency of 9.7 GHz but later resonated at 10.15 GHz after the design.

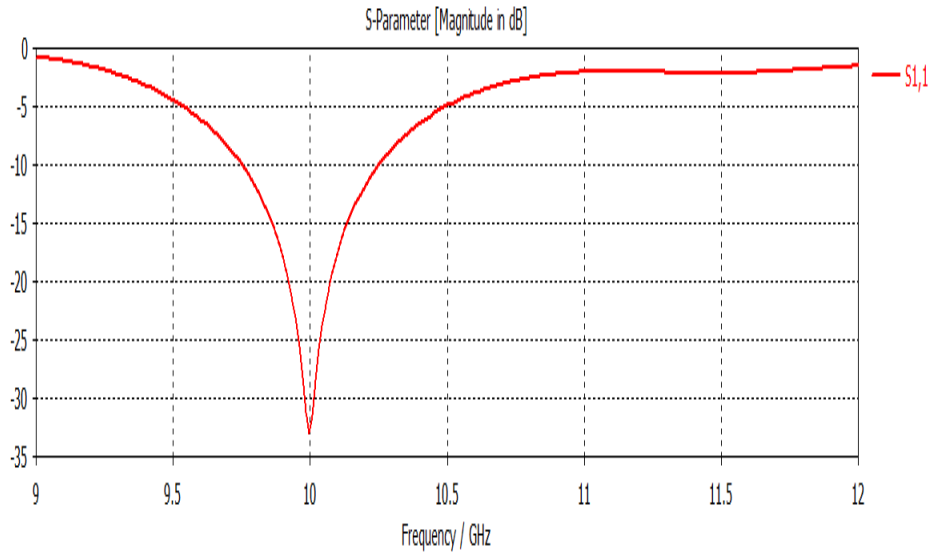

**Figure 5: Return Loss of the Proposed MPA**

**5.2 Voltage Standing Wave Ratio**

Figure 6 shows the VSWR of the proposed MPA design. For mobile communication systems, the value of VSWR should be small (not more than 2.5 dB and close to 1.0 dB) in order to ensure proper impedance matching and little amount of reflected power. The proposed MPA design achieved a VSWR of 1.05 dB at a resonant frequency of 10 GHz. This VSWR performance of the proposed patch antenna has an improvement of 50.70% over the VSWR performance (2.13 dB) of the benchmarked patch antenna design. Another merit of the proposed patch antenna design is that over the selected operational frequency band of 9.7542 GHz – 10.25 GHz, the VSWR stayed below the 2.5 dB baseline. However, the benchmarked patch antenna design exceeded this baseline for most of the operational frequencies in its band of 9.95 GHz to 10.35 GHz.

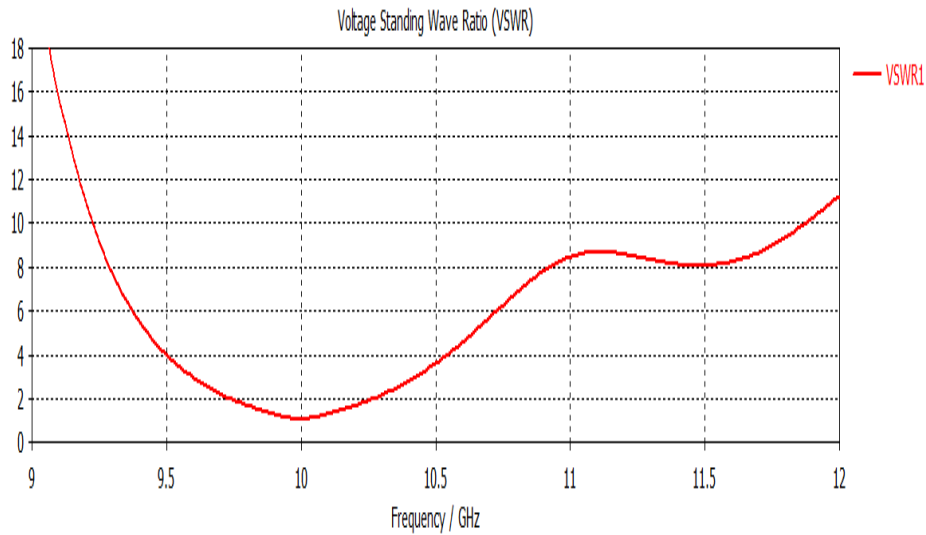

**Figure 6: VSWR of the Proposed MPA**

**5.3 Gain**

Figure 7 shows the 3D gain plot of the proposed MPA design. The proposed patch antenna efficiency is intimately connected to the gain performance. The proposed patch antenna design has a gain of 4.6 dB or 7.2 dBi which is considered acceptable in this scenario of compact antenna design. This gain performance has an improvement of 61.44% over the gain performance (4.46 dBi) of the benchmarked patch antenna design.

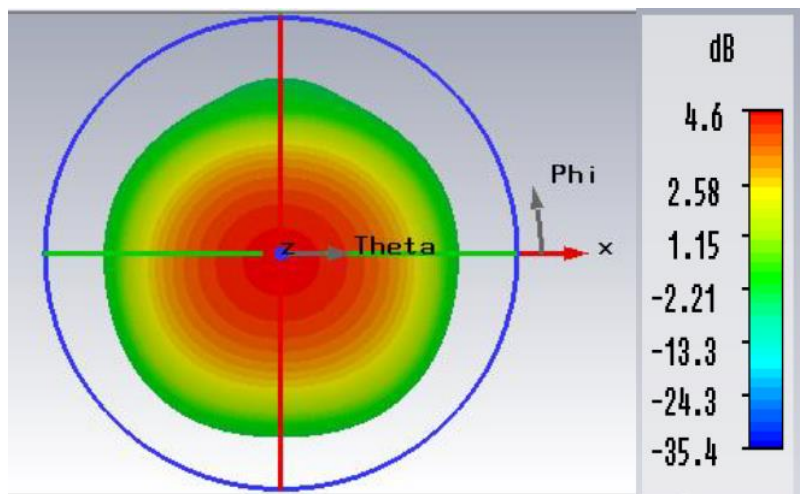

**Figure 7: Gain of the Proposed MPA**

**5.4 Radiation Pattern**

Figure 8 shows the 2D radiation pattern of the proposed MPA design. In comparison with the benchmarked patch antenna design, the proposed design gives a more omnidirectional radiation pattern which makes it suitable for mobile communication systems.

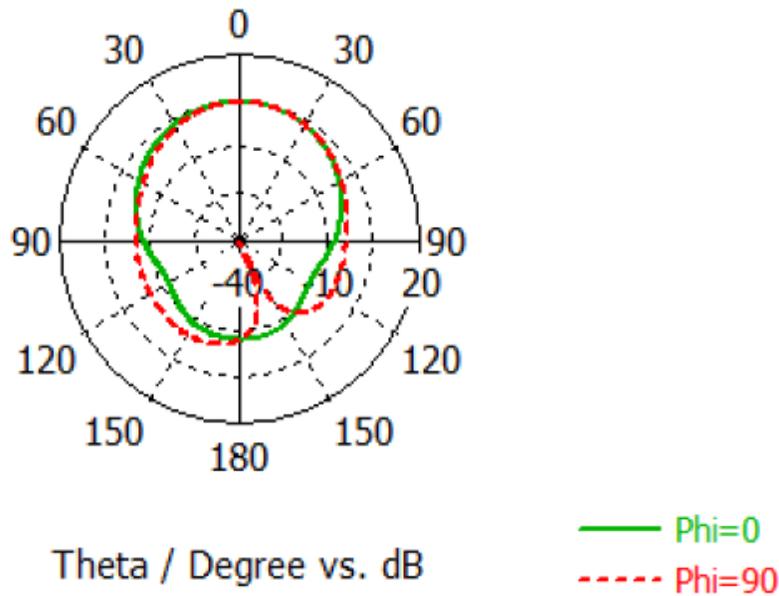

**Figure 8: Radiation Pattern of the Proposed MPA**

The summary of results for the proposed and benchmarked patch antenna design is as shown in Table 2.

**Table 2: Comparison of Results**

| Performance Metrics | Benchmarked Design | Proposed Design | Percentage Improvement |
|---|---|---|---|
| **Return Loss** | -18.27 dB | -31 dB | 69.68% |
| **VSWR** | 2.13 dB | 1.0 dB | 50.70% |
| **Gain** | 4.46 dBi | 7.2 dBi | 61.44% |
| **Patch Area** | 71.4 mm$^2$ | 51.6 mm$^2$ | 38.37% (size reduction) |
| **Bandwidth** | 400 MHz | 500 MHz | 25% |
| **Ground Plane Area** | 400 mm$^2$ | 289 mm$^2$ | 38.41% (size reduction) |
| **Resonant Frequency Offset/Error** | 0.45 GHz | 0 GHz | 100% (no frequency offset error) |

## 6.0 CONCLUSION AND RECOMMENDATIONS

In this research, a compact X-band MPA has been proposed for wireless systems. Over the years, the exponential increase in the demand for mobile data and wireless services has led to a myriad of technological advancements for mobile networks and wireless systems. Inspired with this rapidly-changing technological trends and advancements in the field of mobile communications, this research is proffering an X-band MPA design that will be potentially useful for wireless systems. The proposed patch antenna exhibits a radiation pattern with a gain of 7.2 dBi, resonates at 10 GHz with a return loss of -31 dB and it has a low profile structure of 17 mm × 17 mm × 1.6 mm. The proposed MPA design demonstrates marked performance improvement over a benchmarked patch antenna design. As a possible future research direction, it is recommended that non-contact based feeding method should be applied in order to cater for health and safety issues associated with the specific absorption rate (SAR) of antennas embedded in mobile devices. Other patch shapes can also be applied in order to explore alternative miniaturization techniques. Finally, fabricating the proposed MPA will further substantiate the results obtained from the simulation.